# Global Reduction in Ship-tracks from Sulfur Regulations for Shipping Fuel

Tianle Yuan[1,2], Hua Song[2,3], Robert Wood[4], Chenxi Wang[1,2], Lazaros Oreopoulos[2], Steven Platnick[2], Sophia von Hippel[5], Kerry Meyer[2], Siobhan Light[6], Eric Wilcox[7]

[1]Joint Center for Earth Systems Technology, University of Maryland, Baltimore County
[2]Sciences and Exploration Directorate, Goddard Space Flight Center
[3]SSAI, Inc.
[4]Department of Atmospheric Sciences, University of Washington
[5]University of Arizona
[6]University of Maryland
[7]Desert Research Institute

**Abstract**

**Ship-tracks are produced by ship-emitted aerosols interacting with marine low clouds. Here we apply deep learning models on satellite data to produce the first multi-year global climatology map of ship-tracks. We show that ship-tracks are at the nexus of cloud physics, maritime shipping, and fuel regulation. Our map captures major shipping lanes while missing others, reflecting the influences of background cloud and aerosol properties. Ship-track frequency is more than 10 times higher than expected from a previous survey. Interannual fluctuations in ship-track frequency reflect variations in cross-ocean trade, shipping activity, and fuel regulations. Fuel regulation can alter both detected ship-track density and pattern of shipping routes due to cost economics. The new fuel regulation, together with the COVID-19 pandemic, reduced ship-track frequency in 2020 to its lowest level in recent decades across the globe and may have ushered in a new era of low ship-track frequency. We estimate the aerosol indirect forcing induced by the fuel regulation to be between 0.02 and 0.22 Wm$^{-2}$.**

**Text**

Ship-tracks were first observed in early weather satellites as 'anomalous cloud lines'(*1*) more than 55 years ago. They appear as quasi-linear tracks in marine low cloud fields (Figure 1). Detection of ship-tracks in satellite data relies on a reflectance contrast between background and ship-track clouds (Figure 1). The existence and detection of the contrast depends on various factors such as the wavelength of the observation, and the background properties of cloud and aerosols, tiny suspended airborne particles(*2–5*). As such, even though ship emissions always affect clouds by increasing aerosol concentrations, not all of them produce detectable ship-tracks. Ship-emitted aerosols produce ship-tracks by increasing the concentration of droplets in marine low clouds, which makes them appear brighter(*6, 7*), the so-called aerosol indirect effects. Aerosol indirect effects act as a radiative forcing to the Earth's climate by modifying cloud reflectance and amount and thus affecting the Earth's energy balance(*8, 9*). The aerosol indirect forcing partially counterbalances the radiative forcing caused by greenhouse gases. Existing estimates of aerosol indirect forcing from ship-emitted aerosols range from -0.06Wm$^{-2}$ to -0.6Wm$^{-2}$ (*10–15*). In addition to directly contributing to aerosol indirect forcing, ship-tracks have also been studied to understand aerosol indirect effects in general because they are idealized laboratories where aerosol effects can be clearly separated from effects of meteorology and other factors(*16*). Aerosol indirect forcing as a whole is the leading source of uncertainty in our estimate of various anthropogenic forcing components and its magnitude has significant implications for the Earth's climate sensitivity(*9, 17*). A strongly negative aerosol indirect forcing implies a climate that is highly sensitive to forcing by greenhouse gases as it makes the observed climate change the outcome of a small net positive forcing. In addition, ship-tracks can be viewed as inadvertent cloud brightening geoengineering experiments whose better understanding is necessary to consider deliberate experiments(*4, 15, 18*).

Despite the research interest in ship-tracks, only a single one-year global survey has been carried out so far(*19*). It rejects more than 99% of observations due to data selection criteria(*19*). The lack of comprehensive global sampling of ship-tracks hinders the studies of aerosol indirect effects and geoengineering despite important progress made by analyzing manually labeled samples(*16, 20–24*). Here we combine deep learning models and global satellite observations to automatically identify ship-tracks at unprecedented scales in NASA's Aqua MODerate resolution Imaging Spectrometer (MODIS) daytime data. We show that large-scale sampling of ship-tracks can not only benefit aerosol indirect effect studies, but also reveal unexpected connections among fuel regulation, shipping activity, and ship-tracks. The new dataset will improve progress towards understanding aerosol indirect effects in, and cloud brightening of, marine low clouds.

We train two independent deep neural network models on manually labelled ship-track samples using MODIS 2.1 µm data as input(*3*) (Figure 1). We chose these two models from a pool of candidate models based on their performance and ensemble averaged their results to take advantage of their respective strengths. The ensemble average exhibits better performance than individual model results and generalizes well on test data that are independent of the training data. Details about model performance and validation can be found in the Method section. The two trained models are then applied to Aqua MODIS data between 2003 and 2020.

Figure 2 shows the first global climatological map of ship-track density at 1º resolution. It is the result of processing 0.5 petabytes of data extending from 2003 to 2020. The density is calculated as the number of ship-track pixels divided by the total number of low cloud pixels, i.e., it is the fraction of low clouds belonging to ship-tracks. Low clouds are defined as clouds with top pressure higher than 680 hPa. Overlaid on the ship-track density are emissions of $SO_2$ from global shipping industry(*25*) and MODIS annual mean low cloud fraction during the same period. The pattern of ship-track density generally follows that of ship $SO_2$ emissions, a proxy for shipping routes, in major maritime shipping lanes in the North and Southeast Pacific, in the Southeast Atlantic, in the North Atlantic, and to the South of Australia. The alignment between shipping routes and detected ship-track density varies with region. For example, the two are closely aligned in the Southeast Atlantic and Pacific and to the south of Australia while in the Northern Pacific, especially near the west coast of North America, ship-track density appears much more spread-out than the sharply-defined shipping lines. This reflects the underlying alignment, or lack thereof, between prevailing winds and the direction of shipping lanes(*26*, *27*) and the steadiness of the circulation among other factors. Even when shipping lanes align directionally with detected ship-tracks, there is a consistent shift in location between the two. This is best observed in the Southeast Atlantic and near the Aleutian Islands, where the well-defined ship-track lanes are displaced by about one grid cell (1º in resolution) from the emission lines. This most likely reflects horizontal expansion of ship-tracks with time as they are advected away from their initial formation. Since the number of ship-track pixels increases with expansion, a peak density is expected downwind of the initial formation. The magnitude of shift in the Southeast Atlantic and around Aleutian Islands is quite similar, suggesting similar underlying physics. The shift between emission and peak track density is also apparent for shipping routes with no well-defined lanes such as those off the California coast. Here the detected ship-tracks form a blob instead of distinct lines downwind of shipping lines and their magnitude of shift are larger. Absence of ship-tracks in other regions, e.g. the Tropics, can be explained by unfavorable background cloud and aerosol properties (see Supplemental Material).

An unexpected hot-spot of apparent ship-tracks is found in the Southern Ocean around the South Sandwich Islands where hardly any marine traffic exists. Manual inspection shows that these tracks result from natural volcanic $SO_2$ plumes, which turn into sulfate aerosol plumes modifying cloud properties. Their interaction with clouds and signature in satellite data are almost identical to those of ship-tracks, but at larger scales(*28*, *29*). The agreement between known shipping lanes and detected ship-track density and the independent detection of unexpected volcano tracks are further testaments of the robustness of our method.

Globally, ship-tracks are detected in about 0.3% of marine low clouds at the Aqua MODIS overpass time. Despite the low global value, our ship-track density represents an increase of more than an order of magnitude compared to the sole previous global survey(*19*). The overall low percentage is due to the lack of ship-tracks in the deep Tropics and the fact that ships follow narrow shipping lanes (see shipping emission in Figure 2) leaving most ocean surface free of traffic by large ships. Our global figure represents only a lower bound on the percent of low clouds being affected by ship-emitted aerosols since many impacts are not in the form of ship-tracks(*14*) and explains why large-scale radiative impacts of ship emissions are hard to detect in observations(*30*). Locally, in as much as two to three percent of the low clouds are ship-tracks detected, mainly in subtropical regions dominated by stratocumulus clouds. For example, annual mean ship-track density can exceed three percent in local maxima near the west coast of North America.

In addition to explicitly forming detectable ship-tracks, ship-emitted aerosols can also affect clouds in other ways. For example, as ship-tracks evolve and dissipate with time, they appear as cloudy pixels that are hard to separate from background clouds(*2*, *15*, *22*). Also, when background clouds are polluted, detection of ship-tracks becomes less likely even though ship-emitted aerosols still affect clouds. This is best illustrated using the Southeast Atlantic shipping lane as an example. Few ship-tracks are detected during the fall (see Figure 2) in the segment between 0ºS and 10ºS. Nonetheless, a recent study shows that ship-emitted aerosols have strong effects on cloud droplet size and radiative energy balance for this segment during the same season(*27*). Again, detection of ship-tracks is not a necessary condition for ship-emitted aerosols to be modifying cloud properties(*24*).

Fuel regulations and economic activities are direct drivers of ship-track density. To illustrate the impact of both factors we focus on the Northeast Pacific region because it has the highest ship-track density and there is an Emission Control Area (ECA) under fuel regulations by the International Maritime Organization(*31*)(Outlined in Figure S4). ECAs are established to control the fuel sulfur content of ships that travel inside them. In this ECA, no sulfur standard was implemented before 2005, but fuel sulfur content was limited to 1% in 2010 and 0.1% in 2015. Outside of ECAs, fuel sulfur content was uniformly reduced to 0.5% globally in 2020 from 3.5%(*31*).

The strict 2015 fuel standard inside the ECA has clear impacts on both ship-track density within the ECA and pattern of shipping routes outside of it. Within the ECA, detected ship-tracks density fell by nearly 70% after 2015. On the other hand, the effect of the 2010 standard was much more muted by comparison given the small difference between 2010-2014 and 2005-2009 means. The strong reduction of ship-track density after 2015 is due not only to reduced emissions of sulfur(*32*), which produces a smaller droplet number concentration perturbation and therefore reduces the likelihood of detection, but also to changes in shipping routes(*24*),(*33*). There are two major shipping routes within this region as seen in Figure 2, the northern route connecting Asia to ports of Seattle and Vancouver and the southern route to various ports in California, Mexico and South America. Both routes experience a clear southward shift starting in 2015. We believe both shifts are purposely made by shipping companies to reduce ships' travel time within the ECA(*33*). While the southward shift of the northern route increases the total distance travelled by cruising along the ECA edge to a point that is closest to destination ports and then sailing straight toward them, the time spent within ECA is reduced. The fuel regulation provides enough incentive for shipping companies to make this shift so that ships travel less inside the ECA(*33*). The change in the southern route is even more dramatic. The 2015 fuel policy not only drove ships outside the ECA, but also makes the shipping route more 'contracted'. Before 2015 detected ship-tracks in this area are more 'spread out' forming a blob of high ship-track density. After 2015, ships are much more likely to travel along a narrower corridor. The contraction is apparent as a line of positive density anomaly straddled by strong negative anomalies on both sides.

The impact of the 2020 global fuel standard, representing an 86% reduction of fuel sulfur content outside of ECAs, is the most striking globally. It causes the ship-track density to reach a global minimum in 2020 for both this region and across the globe (Figures 2 and 3). Ship-track density experiences strong reductions in every detected major shipping lane compared to climatology and reaches record lows in the nearly 20-year data record. Except the trans-Pacific and trans-North Atlantic shipping lanes, other shipping lanes are not discernible any longer (Figure S5). Annual mean ship-track density decreases by 50% or more in five major shipping lanes compared to the climatological mean. The decline is even steeper if compared to 2019. Both fuel regulations and temporarily reduced international shipping activity due to the COVID-19 pandemic have contributed to this global reduction(*31*, *34*), although the latter factor is likely a minor contributor. Indeed, ship-based automatic identification system data do not suggest such a strong reduction in annual shipping traffic due to COVID-19[33], and seven months of 2021 data show that detected ship-tracks remain at record-low levels and comparable to 2020 even though the reduction caused by COVID-19 only lasted a few months in 2020. This suggests fuel regulation playing the most dominant role. The global 2020 fuel standard is likely a watershed moment that will permanently reduce the population of detectable ship-tracks.

Fluctuations in international trade and economic activity can also leave fingerprints on detected ship-track density. To illustrate this, we select a region in the Northeast Pacific to capture trans-Pacific shipping activity between Asia and the Americas. The general upward trend of shipping activity between 2003 and 2013 is reflected in the time series of ship-tracks density. The upward trend has a noticeable dip in 2009-2010, coinciding with the aftereffect of the financial crisis of 2008. However, a stronger decrease occurs between 2014-2016, likely caused by a strong slowdown in the Chinese economy. The ship-track density bounces back quickly and reaches another peak in 2018, from where it starts to decrease slightly again, possibly reflecting the trade tension at that time. The density in 2020 drops precipitously due to the new fuel standard and COVID-19 [33].

Changes in fuel regulation offer opportunities to examine how cloud microphysical and optical properties respond to different magnitudes of aerosol perturbations inside ship-tracks. Figure 4 shows perturbations in cloud properties during four periods of different fuel standards. The perturbations are taken as differences in cloud droplet number concentration, $N_d$, and cloud effective radius, $R_e$, between ship-track and background pixels using the MODIS cloud product (*35*) (see Method). We separate the Northeast Pacific into two regions: ECA and its vicinity as non-ECA. Within the ECA region, $N_d$ perturbations, $\Delta N_d$, show a monotonic decrease with increasingly stricter fuel standards, reflecting smaller aerosol perturbations. $\Delta N_d$ decreases from 47 cm$^{-3}$ during 2003-2009 to about 31 cm$^{-3}$ during 2015-2019. The decrease is statistically significant above 95% confidence level. Unexpectedly, the $R_e$ perturbation increases with decreasing $N_d$ perturbation. This is mostly because the background cloud properties have also changed. Clouds of cleaner background are required for ship-tracks to be detected as the amount of emitted aerosols decreases. With lower background $N_d$, the background droplet size is larger. This is in contrast to the neighboring non-ECA area, where neither quantity has a strong trend, except in 2020 when the new global fuel standard takes effect. Cloud optical depth difference between ship-tracks and the background is insensitive to fuel regulations (see Figure 4 and Method). The contrast between ECA and non-ECA regions highlights the impact of the fuel regulation on ship-tracks and the importance of the background clouds.

We estimate the overall impact of the 2020 fuel standard on global ship-emitted aerosols in the context of aerosol indirect effects by examining the droplet number difference between ship-track and background clouds (see Method for more details). We select only clean background clouds with droplet number concentration between 10-20 cm$^{-3}$ for two reasons: first, they occur frequently in the ship-track database; second, they are more likely in an aerosol limited condition and thus a high percentage of ship-emitted aerosols will activate into droplets, which makes changes in droplet number concentration more sensitive to changes in emitted aerosols. We use data from the Northeast Pacific given its abundant ship-track samples. Figure 3C shows a time series of droplet number concentration increase inside ship-tracks. The concentration decreases from a climatological mean of 27 cm$^{-3}$ to 17 cm$^{-3}$ in 2020, or an estimated 37% decrease in ship-emitted aerosols. If we simply scale aerosol emission decrease with indirect forcing due to ship emissions, the 2020 fuel regulation would constitute a positive forcing between 0.02 and 0.22 Wm$^{-2}$.

The first comprehensive survey of global ship-tracks reveals interesting connections between cloud physics, maritime shipping activity, and fuel regulation. Our ship-track database can find more applications in other research areas in the future. For example, the atmospheric chemistry and physics processes taking place between emission of gases to the formation and detection of ship-tracks can be explored to understand the expansion and detection of ship-tracks. Analysis of ship-tracks may be used for fuel regulation compliance in open oceans. Given the order of magnitude increase in detected ship-tracks(*19*), observation-based estimate of radiative forcing from ship emissions needs to be reassessed. In addition, explicitly detected ship-tracks only represent a lower-bound on the ship-emitted aerosols' impact on maritime clouds. For example, a rough estimate based on our ship-track density and previous studies(*19, 27*) would put such forcing on the order of -1Wm$^2$ or more for major ship-track lanes. More in-depth analysis of the aerosol-cloud interactions using comprehensive ship-track data will benefit the understanding of aerosol indirect effects on low clouds in general since such effects are known to be non-linear and sensitive to environmental conditions(*16, 20–24*). Such studies will also help assess the effectiveness and impact of deliberate marine cloud brightening as a geoengineering option. The impact of fuel regulations shown here is a demonstration of both significant human impacts on the marine clouds and our ability to change them with appropriate policies/methods. The evolving impact of the regime changing 2020 fuel standard warrants close monitoring in the coming years given its potential radiative forcing.

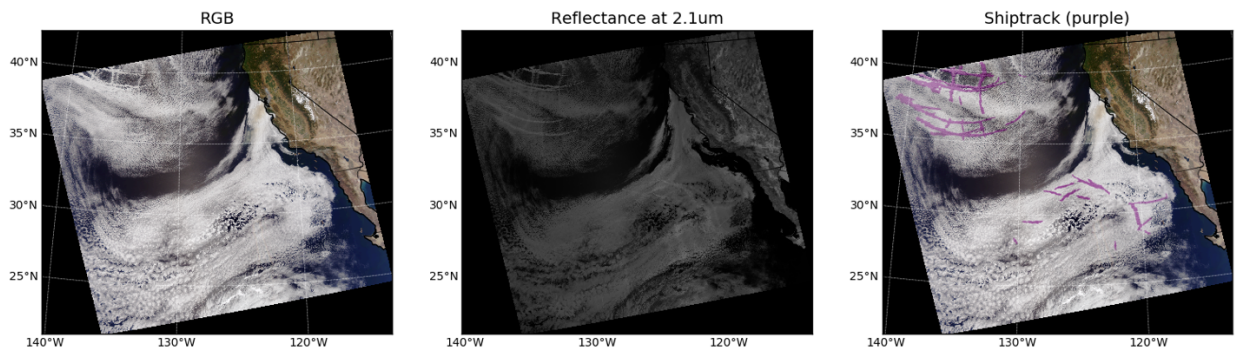

Figure 1: An example MODIS granule off the west coast of North America. Left: true color image; middle: 2.1 μm reflectance; right: detected ship-track mask overlaying on the true color image. 2.1 μm data are better at picking out ship-tracks that are not visible in the true color image.

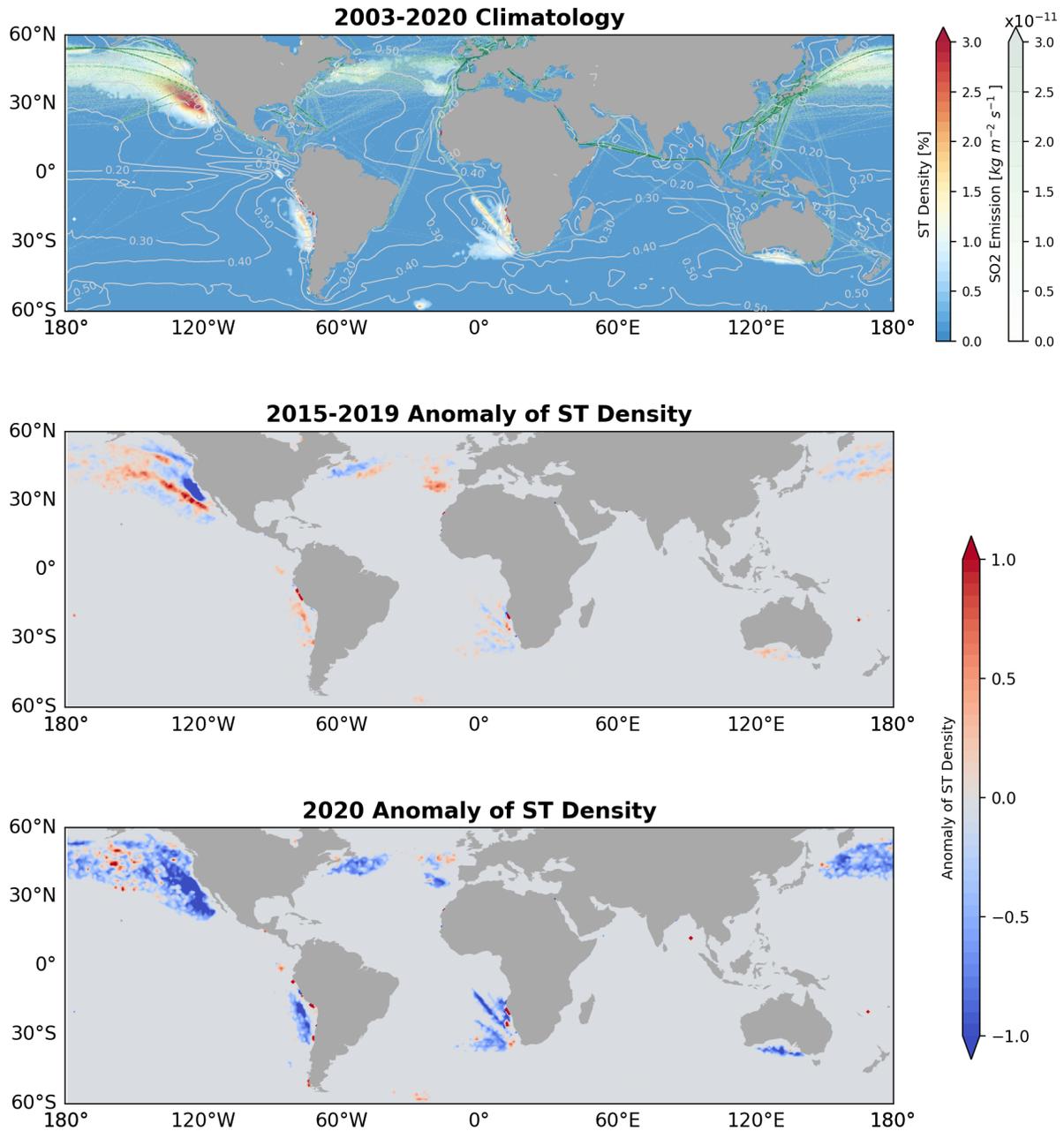

Figure 2: Upper: climatological ship-track density map using data between 2003 and 2020. The gray contour lines show climatology of MODIS low cloud fraction. Green lines are for climatology of ship $SO_2$ emission data. The color maps are for ship-track density. Middle and Lower: Ship-track density anomaly, relative to climatology, maps for four periods: 2015-2019 and 2020. The periods are chosen based on fuel regulation standards. Anomaly maps for other two periods 2003-2009 and 2010-2014 can be found in the supporting materials.

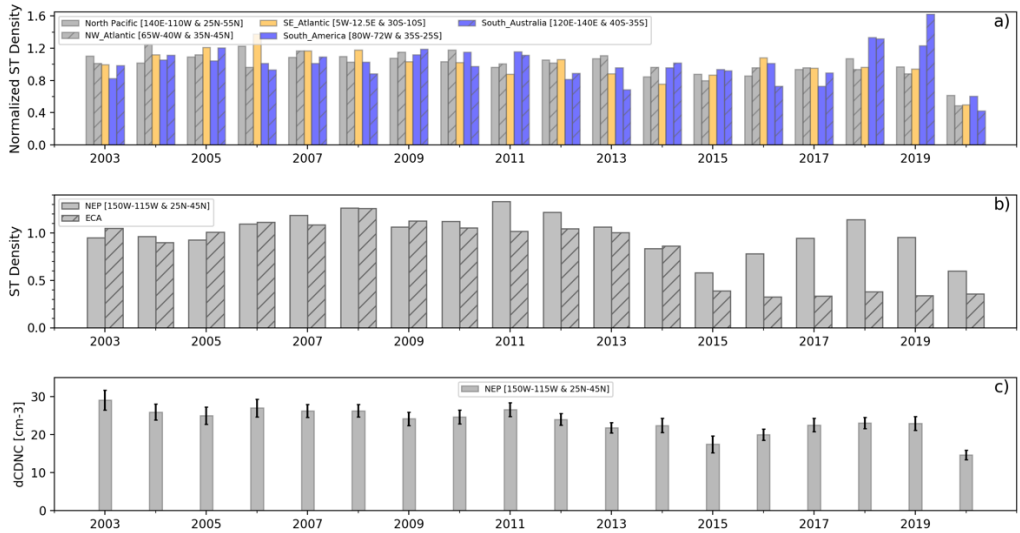

Figure 3. a) normalized time series of ship-track density for five major shipping lanes. Each annual mean data is divided by this region's climatology to make it easier to compare different regions. b) time series in the Northeast Pacific and ECA in particular. Data is not normalized for b. c) time series for ship induced $N_d$ change for clean background clouds.

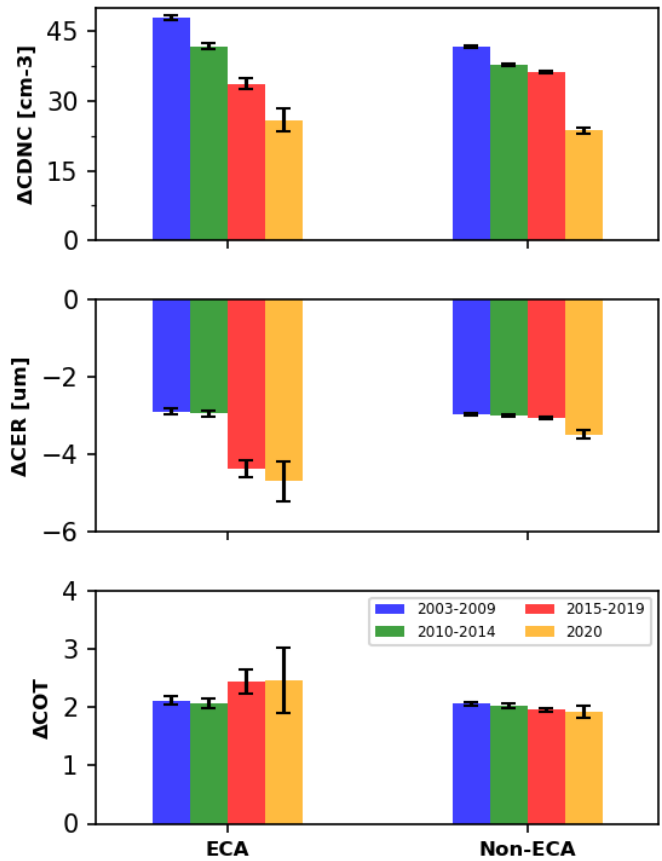

Figure 4: Difference of cloud droplet number concentration, cloud effective radius, and cloud optical depth between ship-track and background for four regions in the Northeast Pacific (see Figure S4 for ECA and non-ECA region shapes).

**Supporting Online Materials:**

a. Ship-track detection

We use a U-net architecture(*36*) to train the model, similar to the night-time model we developed before(*26*). We carry out a set of experiments to adjust hyperparameters such as the number of down sampling blocks and the learning rate. We train the model with different loss functions such as L2 loss, focal loss and cross-entropy. We then pick two models that have complementary strengths as our final ensemble. The output of each model at each pixel can be viewed as a likelihood of that pixel being a ship-track pixel. We simply average two models' output as the final output and use 0.3 as the threshold for binary classification for each pixel. A few examples are given in Figure S1. Our ensemble model achieves an F1 score of 0.81, precision of 0.87, and recall of 0.77 on test data.

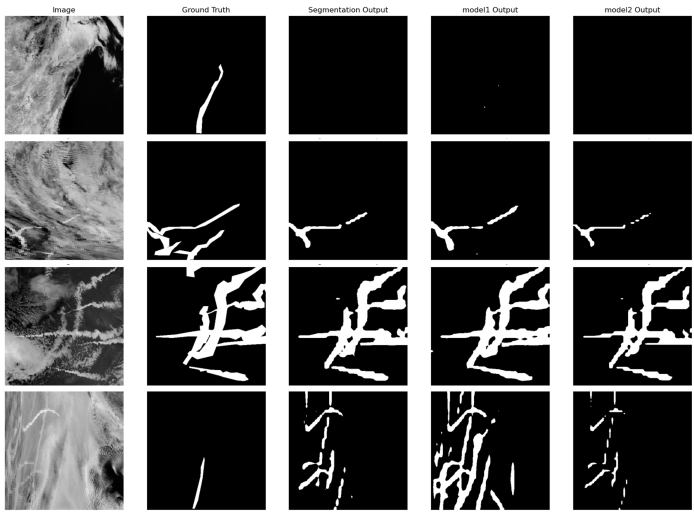

Figure S1: Four kinds of examples: (left)MODIS 2.1um images, (second to the left) manually labelled ship-track masks (white pixels), (middle) ship-track masks from an ensemble of models, (right two) ship-track masks from two models. In the first example, the manual ship-track label is a false positive. In the second example, the models pick out most manual labels and leave out the questionable ones. Model and manual labels agree well in the third, quite complex example. In the last example, the labeler clearly missed many true positives while the models correctly detected them. Most training and test samples fall into second and third kinds.

b. Validation

In addition to validate model outputs on test manual labels, we also independently test our model output against never-seen samples using a trajectory model. This is because the ground truth for ship-tracks is hard to find for satellite data-based detection since in-situ observations are extremely limited. We can only rely on human inspection of the original reflectance or BTD data, which is still not direct validation(*37*). We use Automatic Identification System (AIS) and MERRA-2 reanalysis data to validate our detections. AIS data report real-time ship data such as location and speed every 6 minutes and are publicly available for ships around US coastal regions. We developed a forward trajectory model to facilitate validating ship-track detection. The forward trajectory model uses MERRA2(*38*) near-surface wind to advect ship emission based on AIS data. The 1-hourly averaged 50-meter U and V wind components from MERRA-2 high-spatial resolution (0.625°×0.5°) data are used. The 6-minute AIS ship location data are down-sampled to the half-hourly. Each ship releases a virtual emission parcel at every time step and the forward trajectory model predicts their locations at satellite passing time. We then obtain an expected ship emission track at the MODIS overpassing time by connecting predicted locations of each virtual parcel. These expected tracks are compared with the actual detected ship-tracks for validation.

Figure S2 gives an example where blue lines show actual trajectories of 16 vessels between 00:00 to 22:00 UTC, July 17 2018. Red lines are predicted ship emission tracks at 21:30 UTC using the forward trajectory model. Detected ship-track masks by our algorithms are shown in green. The predicted tracks by forward model match well with detected ship-track masks by our algorithms. We note that ship-track masks can be broken due to broken low clouds and/or overlapping high clouds. Ship number 7's predicted track is slightly off from the actual ship-track mask, possibly reflecting imperfect wind fields or forward model. Overall, the match-up provides an excellent qualitative validation for our detection model.

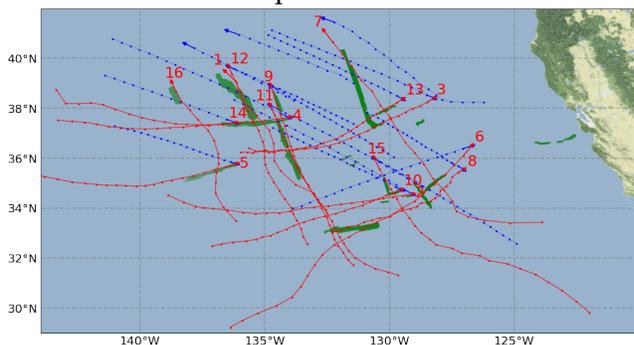

Figure S2: An example of predicted tracks based on wind and AIS data, together with detected ship-track masks from our algorithm. We picked ships (indicated by numbers) whose predicted tracks matched with ship-track masks. Most match really well with ship #7 slightly off, which could be attributed to imperfection in wind field or trajectory model.

c. Analysis of aerosol indirect effects using ship-tracks

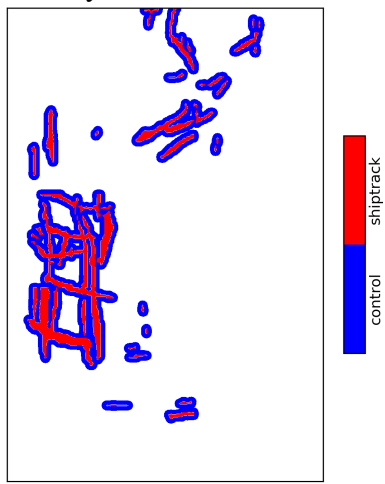

Figure S3: Detected ship track pixels and their background pixels for an Aqua granule taken at 2155 UTC on July 15, 2018. The width of the background (blue color in the figure) is 20 pixels.

For each detected ship-track, we automatically find background pixels surrounding it whose width is 20 pixels on both sides as shown in Figure S3. Each granule, approximately 2030x1350, is then broken into small blocks of size 128x128. Within each block, we calculate mean values of cloud variables such as droplet effective radius, cloud optical depth, and cloud droplet number concentration for the background and ship-track pixels. The difference between background and ship-track pixels' means is taken as the impact caused by ship-emitted aerosols. Aggregating values from many such blocks leads to the mean values and 95% uncertainty ranges reported in Figure 4.

d. <u>Estimating Aerosol Indirect Forcing due to regulation</u>

First, we have $\frac{\Delta COD}{COD} = \frac{\Delta LWP}{LWP} - \frac{\Delta R_e}{R_e}$, where COD is cloud optical depth, LWP is liquid water path, and $R_e$ is cloud effective radius(21). $\Delta COD$ is thus usually proportional to background COD if the ratio is close to constant, which is mostly supported by our data (Figure 4).

Without considering aerosol effects on cloud fractions, cloud albedo sensitivity to aerosols can be taken as the sum of the Twomey effect and aerosol induced LWP changes:

$$S = \frac{dA_c}{dN_d} = \frac{A_c(1-A_c)}{3N_d} \times (1 + \frac{5}{2}\frac{d\ln\ln LWP}{d\ln N_d})$$

where S is the susceptibility of cloud albedo ($A_c$) to droplet number concentration ($N_d$)(23). Aerosol indirect forcing is therefore:

$$\Delta SW_{TOA} = - SW_{downwelling} \times CF_{liquid} \times A_c \times (1 - A_c) \times (\frac{1}{3} + \frac{5}{6}\frac{d\ln\ln LWP}{d\ln N_d}) \times \Delta \ln N_d.$$

We have $\Delta SW_{TOA} \sim \Delta \ln N_d$ and $\Delta \ln N_d \approx \frac{\Delta N_d}{N_d}$. To get $\Delta \ln N_d \approx \frac{\Delta N_d}{N_d}$, we target background clouds that are clean. Clean background clouds ensures that high fraction of ship-emitted aerosols would activate into droplets and the difference between ship-track and background clouds in $N_d$, i.e. $\Delta N_d$, can therefore better capture the changes in the number of ship-emitted aerosols(39). We find these clean clouds in each year and time series of $\Delta N_d$ for such clean clouds reflects the impact of fuel regulations on $N_d$ (32). Once we estimate the impact of fuel regulations on $\Delta N_d$ using clean background clouds, we assume the magnitude of change in underlying ship-emitted aerosols due to fuel regulation is the same for all ships and conditions on average. The ship emission change due to regulation should have no correlation with background cloud properties, which we believe is a reasonable assumption. We can thus obtain $\Delta \ln N_d \approx \frac{\Delta N_d}{N_d}$.

From the literature, aerosol indirect forcing due to ship emissions is estimated to be -0.06Wm$^{-2}$ to -0.6Wm$^{-2}$ and therefore forcing due to fuel regulation is 37% of that as shown in the text, i.e., 0.02Wm$^{-2}$ to 0.22Wm$^{-2}$, assuming the LWP and cloud fraction responses are constant.

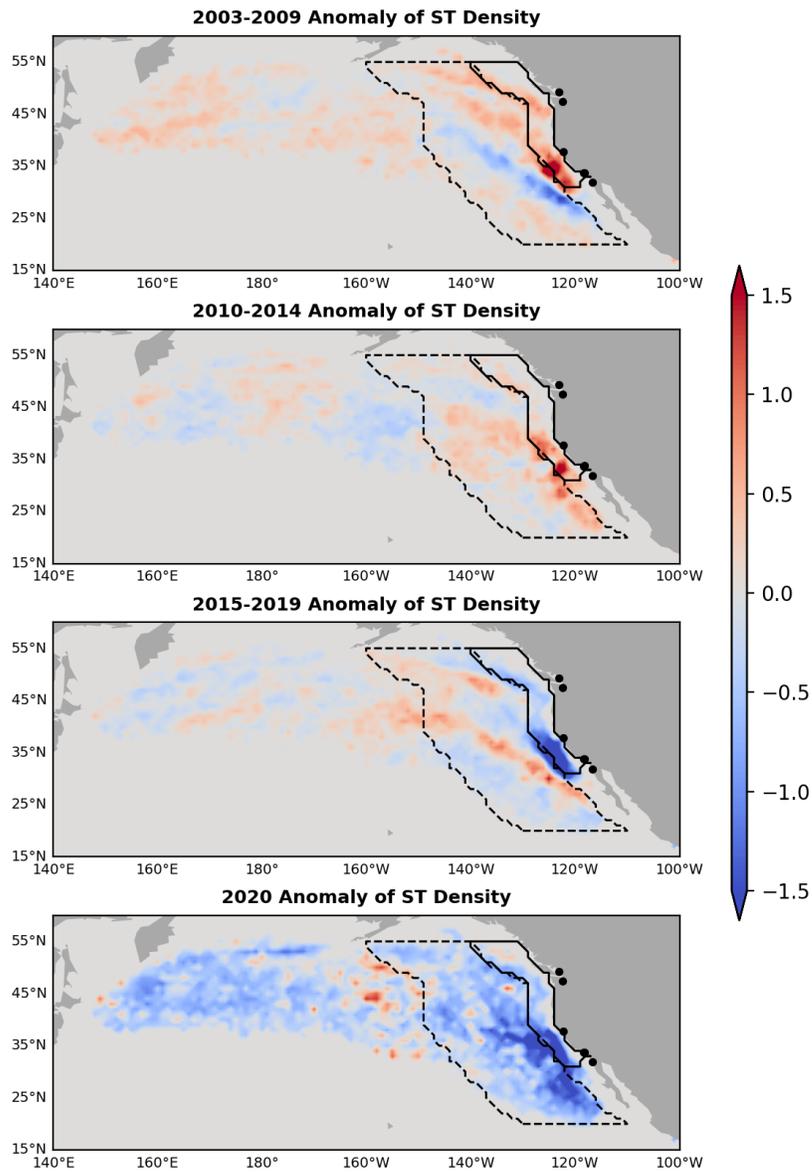

Figure S4: Focusing on the Northern Pacific region. Solid green and black lines outline the North and South ECAs in the text. Corresponding dashed lines are neighboring non-ECA regions.

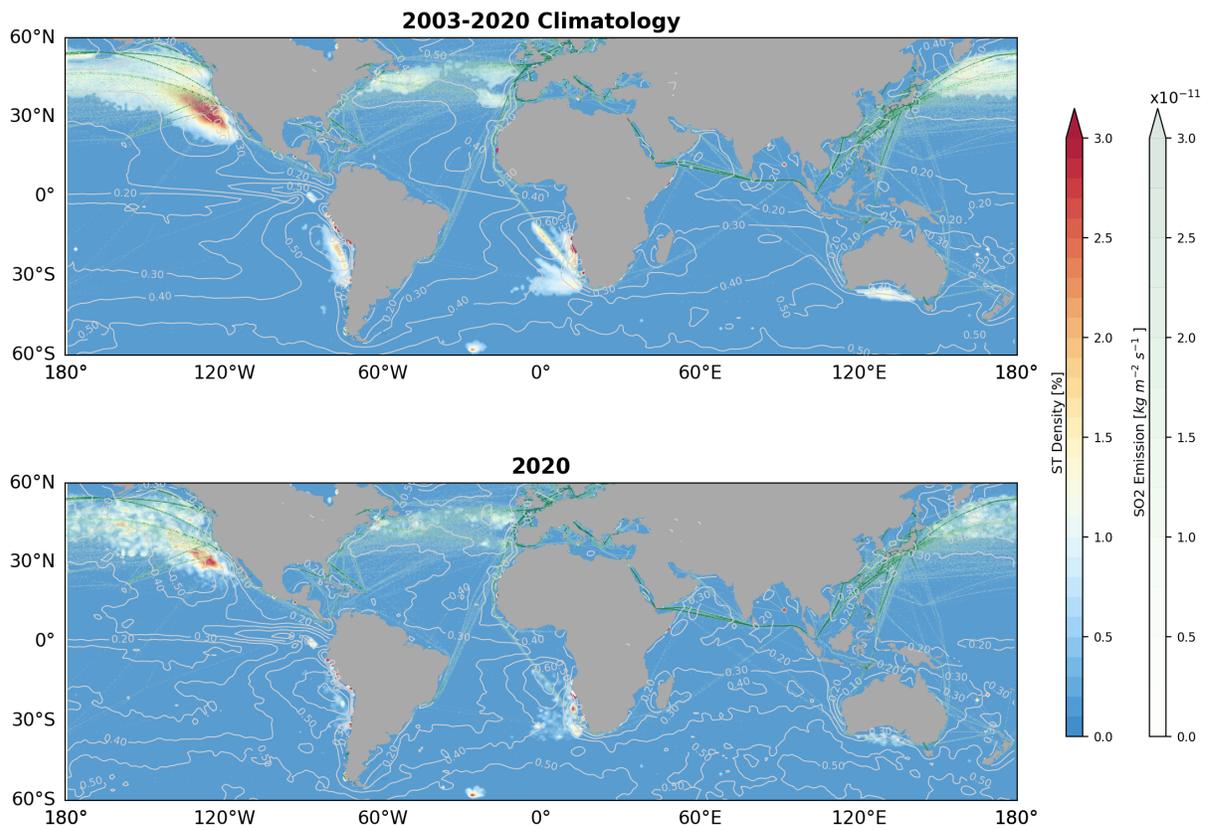

Figure S5: Comparing ship-track density maps of 2020 and 2003-2020 climatology. The southern hemisphere ship lanes have nearly disappeared due to the fuel regulation and temporary covid slump in shipping activity.

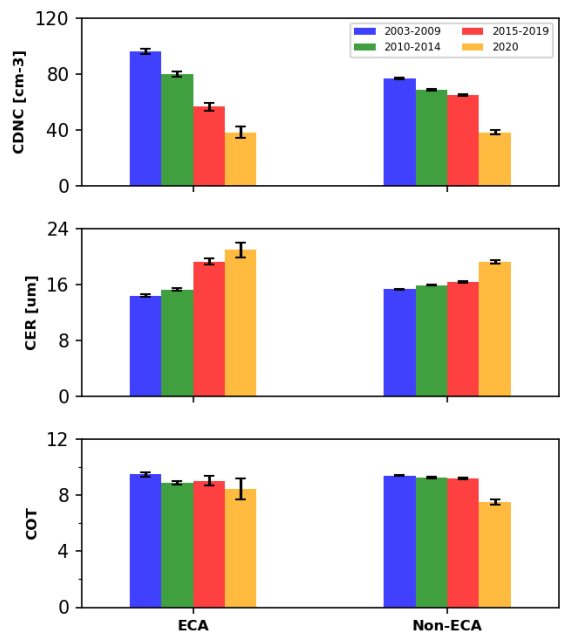

Figure S6: Background cloud properties, cloud effective radius (CER), cloud optical depth (COT), and cloud droplet number concentration (CDNC) for ECA and non-ECA regions in Figure 4.

Impact of cloud types and background pollution on ship-track density

Cloud types also affect both the formation and detection of ship-tracks. The good alignment between shipping lane and detected ship-track density is only found in the stratocumulus dominated regions. In the tropics, few ship-tracks are detected even along busy shipping lanes because scattered cumulus clouds dominate there. Cumulus cloud fields in the tropics are highly scattered with low cloud fraction, occasional presence of ice phase, and often overlap with high cirrus and thick high clouds, none of which are favorable for ship-track detection and contribute to absence of ship-tracks in the tropics. Deeper boundary layers in the tropics are also unfavorable for ship-track formation due to dispersion of ship-emitted aerosols over larger volumes and therefore smaller perturbation in aerosols. Similar dependence of ship-track formation/detectability on cloud types has been seen in large-eddy simulations(*5*). In broken clouds, tracks may be difficult to detect even if there are significant cloud albedo increases from aerosol emissions. The dependence on cloud types is further supported by analysis of the seasonal cycle. The seasonal cycles of both ship-track numbers and density correlate well with that of low cloud fraction in regions with significant ship traffic (Table S1). Since there is little seasonal variation in ship SO2 emission for these regions according to the emission data(*25*), we can assume that ship traffic is constant with seasons. The positive correlation between ship-track density and low cloud fraction likely suggests influence of cloud type on ship-track formation/detection since the density is already normalized by cloud fraction.

In addition to cloud types, background aerosol concentrations can also affect the detection of ship-tracks. The droplet number concentration in clouds formed in more polluted environment is less limited by the aerosol concentration than by the in-cloud supersaturation. Additional ship-emitted aerosols in such environments have less impact on droplet number concentration with other conditions being equal, resulting smaller contrast in 2.1 µm reflectance and suppressing detection of ship-tracks. This can be illustrated in the Southeast Atlantic. The busiest shipping lane here connects Cape Town to Europe (see emission in Figure 2). While the number of ships in this lane remains constant, there is significant contrast in ship-track density between its northern and southern segments. We found a significant (r = -0.49, p <0.01) anti-correlation between the north-south differences in the background sulfate concentration of the boundary layer from reanalysis(*38*) and the ship-track density. In other words, with equal number of ships traversing the shipping lane, higher background aerosol concentration tends to reduce the number of detected ship-tracks.

| Region | NE Pacific | NW Atlantic | SE Atlantic | South America | NW Pacific |
|--------|-----------|-------------|-------------|---------------|------------|
| r      | 0.6       | 0.87        | 0.85        | 0.85          | 0.81       |

Table S1: Correlation between seasonal cycles of ship-track density and low cloud fraction.

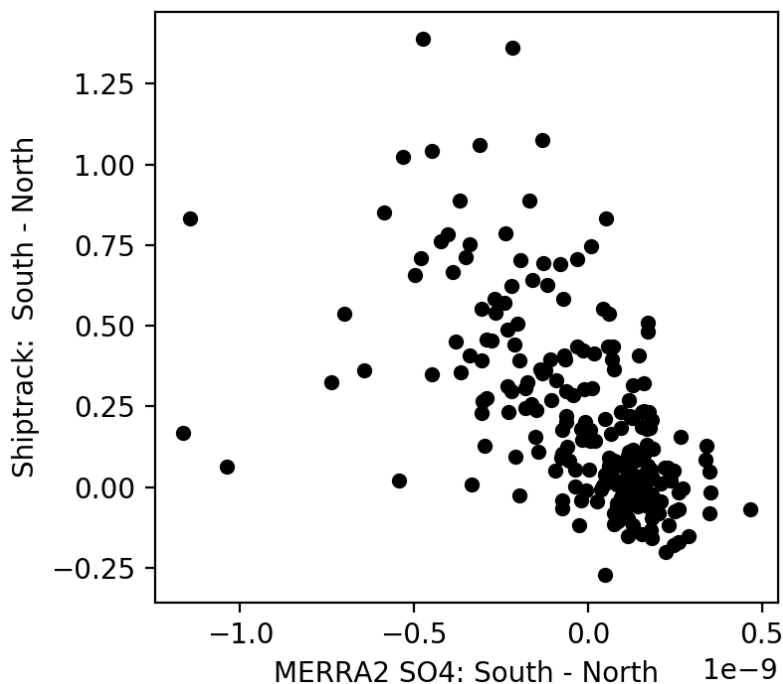

Figure S7: correlation between south-north differences in boundary layer sulfate concentration and ship-track density. The north-south difference for both quantities is calculated for each month and this plot contains 216 months of data. The regions are defined by 2W-10E, 25S-10S and 15W-2W, 10S-0S.